\documentclass[pdflatex,sn-mathphys-num,iicol]{sn-jnl}
\usepackage[utf8]{inputenc}
\usepackage{cuted}
\usepackage{graphicx}%
\usepackage{latexsym}
\usepackage{amsmath,amssymb,amsfonts}%
\usepackage{amsthm}%
\usepackage{mathrsfs}%
\usepackage[title]{appendix}%
\usepackage{textcomp}%
\usepackage{manyfoot}%
\usepackage{booktabs}%
\usepackage{algorithm}%
\usepackage{algorithmicx}%
\usepackage{algpseudocode}%
\usepackage{listings}%
\theoremstyle{thmstyleone}%
\usepackage{todonotes}
\theoremstyle{thmstyletwo}%
\theoremstyle{thmstylethree}%
\usepackage{textcomp}
\usepackage{rotating}
\usepackage{bm}
\usepackage{float}
\usepackage{etoolbox}
\PassOptionsToPackage{colorlinks=true,linkcolor=blue,citecolor=blue, urlcolor=blue,anchorcolor = blue}{hyperref}
\usepackage[colorlinks=true,linkcolor=blue,citecolor=blue, urlcolor=blue,anchorcolor = blue]{hyperref}
\usepackage{epsf}
\usepackage{graphics}
\usepackage[export]{adjustbox}
\usepackage{tabularx}
\usepackage{multirow}
\usepackage{array} 
\usepackage{makecell}  
\usepackage{diagbox}  
\usepackage{colortbl}  
\usepackage{diagbox}
\usepackage{slashbox}
\usepackage{epsfig,latexsym}
\usepackage[section]{placeins}
\usepackage{booktabs}
\usepackage{changes}

\def\bea{\begin{eqnarray}}
\def\eea{\end{eqnarray}}

\begin{document}

\title{Entropy-Based Analysis of Urban Pollutant-Weather Correlations}

\author[1]{\fnm{Suchismita} \sur{Banerjee}} 

\author[2]{\fnm{Koyena} \sur{Ghosh}} 

\author[2]{\fnm{Moumita} \sur{De}} 

\author*[1]{\fnm{Urna} \sur{Basu}}\email{urna@bose.res.in}

\author[3]{\fnm{Banasri} \sur{Basu}} 

\affil[1]{\orgname{S. N. Bose National Centre for Basic Sciences}, \orgaddress{\city{Kolkata}, \postcode{700106}, \country{India}}}

\affil[2]{\orgname{Maulana Abul Kalam Azad University of Technology}, \orgaddress{\state{West Bengal}, \postcode{741249} \country{India}}}

\affil[3]{\orgname{Indian Statistical Institute}, \orgaddress{\city{Kolkata}, \postcode{700108}, \country{India}}}

\abstract{We employ statistical physics and information-theoretic methods to quantify the dependencies between key atmospheric pollutants and meteorological variables across multiple Indian cities. To capture both linear and nonlinear relationships, we introduce a Composite Correlation Index (CCI) that combines the Pearson correlation coefficient with entropy-based measures, including mutual information and conditional entropy.
Based on the CCI values, cities are clustered into distinct groups, uncovering regional similarities in pollutant–meteorology interactions that may reflect shared climatic or environmental conditions.  To explore temporal structure and causal dynamics, we analyze the relationship between particulate matter (PM$_\text{2.5}$) and relative humidity (RH) using transfer entropy, which reveals a bidirectional flow of information in most locations. Further time-domain analysis via time-delayed mutual information shows that, in many cities, the dependence between PM$_{2.5}$ and RH peaks at zero lag and decays exponentially thereafter, indicating predominantly contemporaneous interactions with limited memory. This integrative framework provides a robust approach to characterizing atmospheric interaction regimes, bridging statistical physics with environmental complexity and revealing new insights into the pollutant-meteorology dynamics.}


\maketitle


\section{Introduction} 
Air pollution dynamics can be viewed as a classic example of an open, driven-dissipative  system~\cite{Chaos_2010,Bak_2007,Nic_1977,Love_2015,Love_2013}  where energy and matter continuously flow through chemical, physical, and weather-related processes. 
These far-from-equilibrium dynamics gives rise to  complex spatiotemporal patterns~\cite{Run_2019, Sein_2016, Jac_1999}, shaped by both natural and anthropogenic factors. From a physics perspective, such systems challenge classical equilibrium models and call for tools from nonlinear dynamics~\cite{Raga_1996}, stochastic processes~\cite{Hwa_2013}, and information theory~\cite{Gol_2022}.

Pollutants (such as PM$_{2.5}$, PM$_{10}$, SO$_{2}$ and NO$_{2}$ and many more) are not merely passive tracers; they undergo nonlinear transformations~\cite{Bin_2019}, interact through feedback mechanisms~\cite{Im_2022}, and are transported via advection-diffusion processes~\cite{Lin_2021}, all modulated  by meteorological (such as temperature,  precipitation etc.) forcing.  Understanding  the interdependencies between  pollutants and weather variables is essential not only for accurate predictive modeling and air quality assessment, but also for uncovering emergent behavior in complex environmental systems~\cite{Cov_1999,San_2002}.

This study investigates the statistical structure and directional dependencies among key atmospheric pollutants and meteorological drivers. While prior studies have shown that weather significantly affects pollutant dispersion and accumulation~\cite{Row_2024,Nak_2025,Mal_2023}, conventional statistical models often rely on linear assumptions and overlook both non-stationarity~\cite{Garsa_2023} and directional causality. To address these limitations, we adopt an entropy based information-theoretic approach.

Information theory, rooted in statistical physics and complex system science, provides powerful tools for characterizing structure and dynamics in multivariate systems~\cite{Gol_2022,Andrea_2023}. Metrics such as Shannon entropy~\cite{Shanon}, Mutual Information (MI)~\cite{Fras_1986,Steu_2002}, and Transfer Entropy (TE)~\cite{TE_2000} have proven valuable across fields from neuroscience to climate science~\cite{Lel_2015,Cohen_2017}, and here we extend their application to urban air quality dynamics.

Using a multi-year dataset covering multiple Indian cities, we analyze interactions between pollutants ($\text{PM}_{2.5}$, $\text{PM}_{10}$, $\text{NO}_{2}$, and $\text{SO}_{2}$) and meteorological variables (relative humidity RH and ambient temperature AT). We compute the standard Pearson correlation coefficient to estimate the  linear correlations and use entropy-based measures to quantify uncertainty, dependency, and the directional flow of information~\cite{Pompe_2011,Granger_2009} thereby capturing  both linear and non linear relationships. Specifically, Pearson correlation, Mutual Information and relative conditional entropy are used to assess static relationships, while Transfer Entropy estimates  temporal causality. Additionally, Lagged MI is employed  to investigate  the decay of dependency over time.

Our results reveal  spatial heterogeneity as well as similarities  in pollutant–meteorological couplings, suggesting that regional climate and geography play a fundamental role in shaping these dynamics. We find cities cluster into distinct groups with varying levels of interdependence. 
Additionally, transfer entropy analysis, applied to the PM$_{2.5}$–relative humidity (RH) pair provides  evidence of bidirectional information flow in most  cities. Lagged mutual information analysis further reveals that, for some cities, dependencies between PM$_{2.5}$ and RH, peak at zero lag and decay rapidly with increasing lag.

The remainder of the paper is organized as follows: Sec.~\ref{sec:data} describes the dataset and spatial–temporal coverage. Section~\ref{sec:prelim} presents preliminary  analyses and visualization of temporal trends. Section~\ref{sec:corr} focuses on the methodology and main results, including entropy-based metrics and their interpretation.
The temporal characteristics of   PM\textsubscript{2.5} and RH is discussed in Sec.~\ref{sec:temp}.
We conclude with some general remarks in Sec.~\ref{sec:concl}. 

\section{Data Description} \label{sec:data}
\subsection{Location, Source and Collection Period}
The data used in this study is obtained from the Central Pollution Control Board (CPCB) of India~\cite{1}, which provides publicly available, quality-assured environmental monitoring data. The dataset comprises daily average concentrations of various atmospheric pollutants and meteorological variables collected from various monitoring stations ($N_m$), that recorded the required set of variables in each location across multiple cities [summary given in Table~\ref{tab:1}]. 
To be precise, data is collected from $87$ monitoring stations, across $24$ cities~\cite{2}, including seven Tier-1 cities (characterized by high population density, robust infrastructure and economic activity) 
and seventeen Tier-2/3 cities (smaller urban centres with growing population and developing infrastructure). 
The chosen cities represent a diverse range of geographic locations~\cite{3}, climatic zones~\cite{4}, population densities~\cite{5}, industrial profiles, and emission sources, making them suitable for a comprehensive comparative study of air quality dynamics and their interactions with meteorological variables. \\

\noindent{\textbf{Selected Environmental Parameters}}: In our analysis, we  focus on a set of key variables that include both air pollutants and meteorological factors. Specifically, we select \textbf{$\text{PM}_{2.5}$} (particulate matter with diameter $< 2.5 \,\mu m$), \textbf{$\text{PM}_{10}$} (particulate matter with diameter $< 10 \, \mu m$), \textbf{$\text{NO}_{2}$} (nitrogen dioxide) and \textbf{$\text{SO}_{2}$} (sulfur dioxide) due to their known impact on air quality and human health. Among these, \textbf{PM$_{2.5}$} and \textbf{PM$_{10}$} are categorized as particulate (matter-type) pollutants, while \textbf{NO$_2$} and \textbf{SO$_2$} are gaseous pollutants. In addition, two meteorological parameters—\textbf{RH} (relative humidity, expressed as a percentage, representing the amount of moisture in the air relative to the maximum it can hold at a given temperature) and \textbf{AT} (ambient temperature, measured in degrees Celsius, indicating the surrounding air temperature) are included, given their influence on pollutant dispersion and chemical transformation in the atmosphere. The choice of these variables is guided by both data availability and their relevance to the study objectives.
 
\begin{figure}[t]
    \centering        \includegraphics[width=\linewidth]{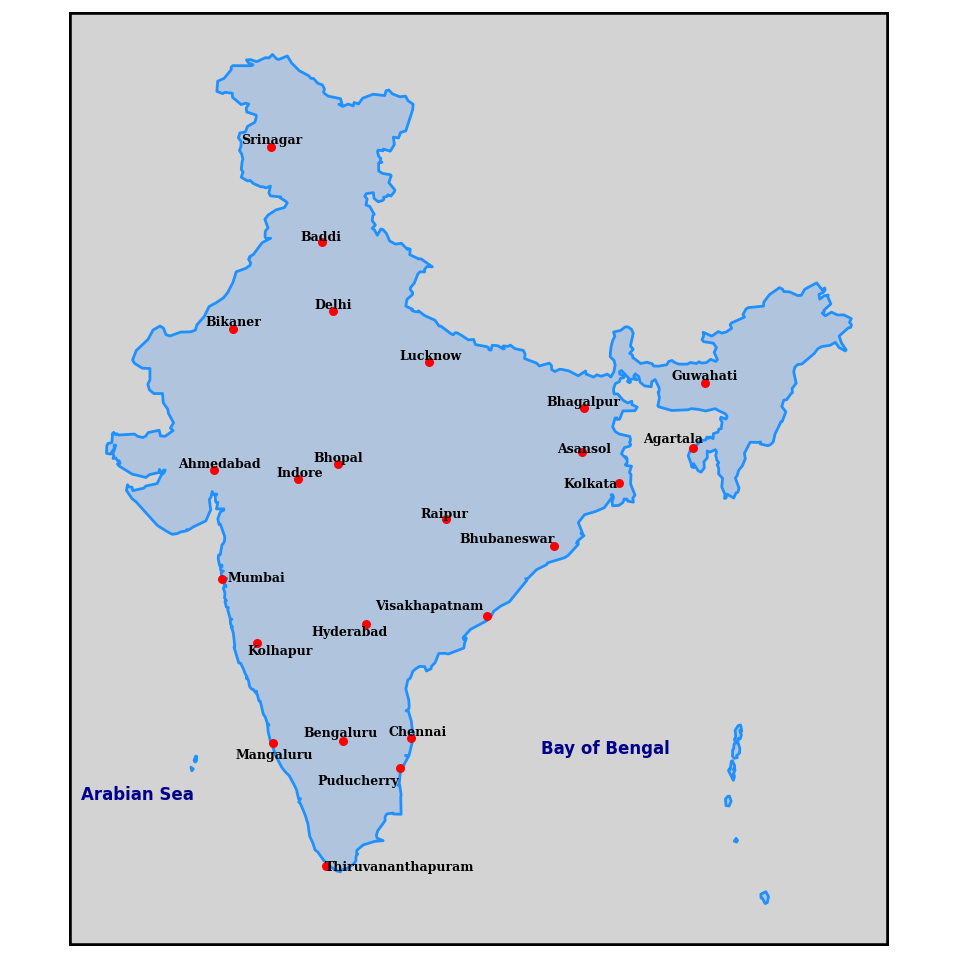}
    \caption{Map of India showing the locations of the selected cities.}
    \label{fig:map}
\end{figure}

\subsection{Data Pre-processing }
Environmental data often suffer from missing values and sensor noise due to hardware failures, power outages, or maintenance activities. To address this issue and to smooth the time series for reliable analysis, we employ Kalman filtering~\cite{Kalman,becker2024kalman}, a recursive optimal estimation technique widely used for noisy time series data.
We apply the Kalman filter separately to each time series (for all the pollutants and meteorological parameters) for each city. The filter not only interpolates missing values but also de-noises the observed signals, preserving the underlying dynamics critical for entropy and information-theoretic analyses.

\section{Preliminary Analysis}\label{sec:prelim}
This study employs a multi-layered, entropy-based approach to investigate the interdependencies and directional relationships between air pollutants and meteorological variables in Indian cities.
As a starting point, a preliminary analysis is conducted to explore the complex temporal behavior of key pollutants and meteorological parameters. This includes generating time series plots and applying non-parametric regression techniques to identify underlying trends and patterns across different regions.

\begin{figure*}[tbh]
    \centering    \includegraphics[width=7.5 cm]{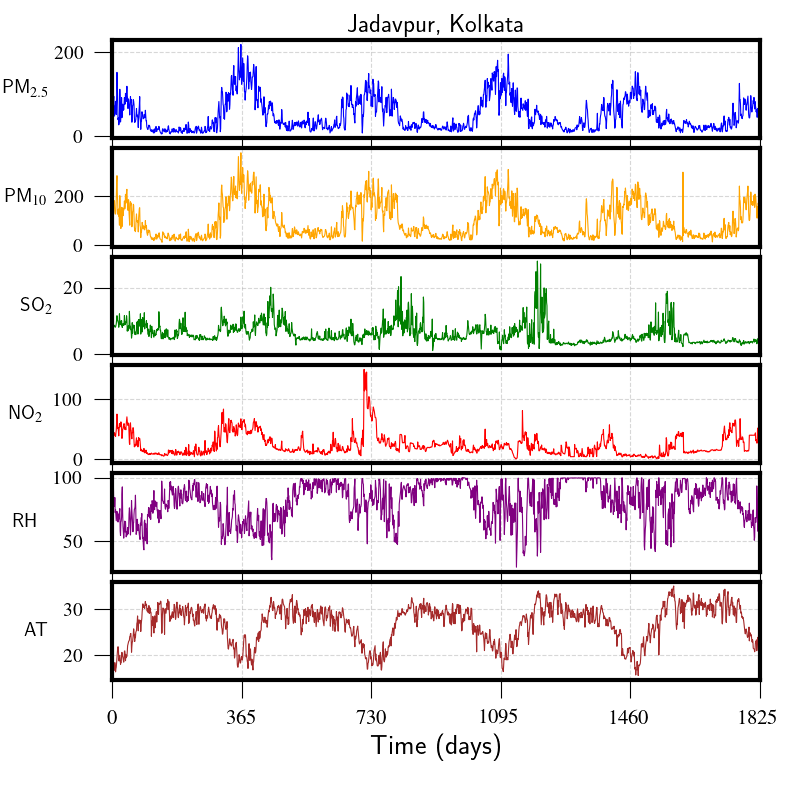}    \includegraphics[width=7.5 cm]{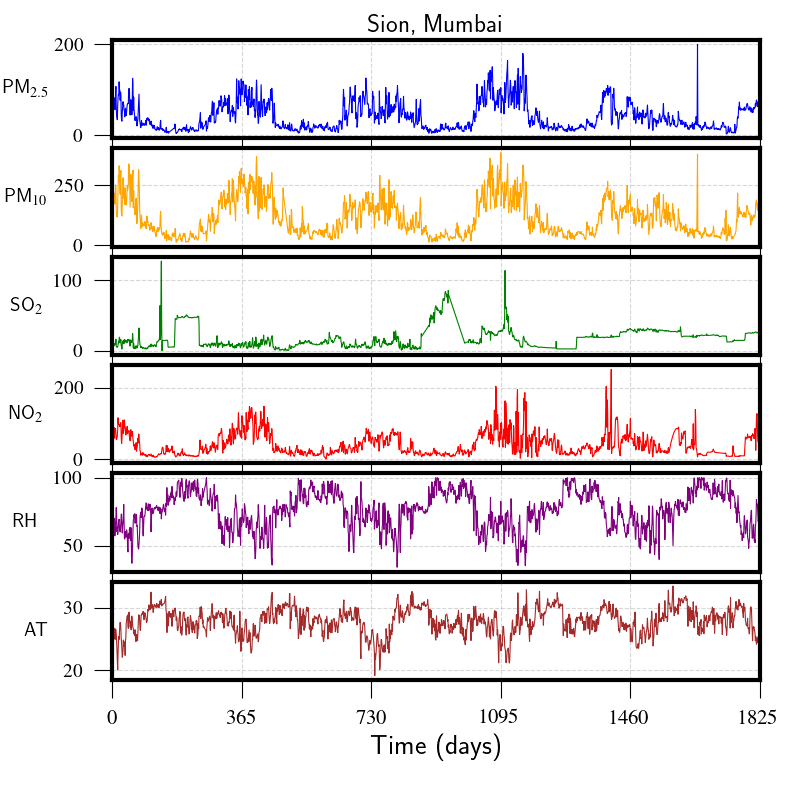}
    \caption{Time-series plots for concentration of the pollutants and the meteorological parameters for two representative stations in Kolkata (left panel) and Mumbai (right panel) for the period Jan, 2020 to Dec, 2024.}
    \label{fig:time_series}
\end{figure*}

\subsection{Temporal Trends}
To visually examine temporal fluctuations and possible seasonal patterns, time series plots of the  pollutant concentrations  and meteorological variables are generated for each location. Fig.~\ref{fig:time_series} depicts such time series plots for two representative stations, namely, Jadavpur in Kolkata and Sion in Mumbai. 

Each panel stacks five years of daily data for six variables - PM\textsubscript{2.5}, PM\textsubscript{10}, SO\textsubscript{2}, NO\textsubscript{2}, RH and AT, so that  seasonal rhythms can be compared at a glance. Both PM\textsubscript{2.5}, PM\textsubscript{10}, surge each winter reaching their peaks around Dec-Jan, 
and decrease  during the warm,  humid monsoon months pointing to strong seasonality and common sources.
Both the weather variables RH and AT show an almost reverse trend, they reach their minimum during Dec-Jan and become maximum in the summer and monsoon months. 
SO\textsubscript{2} remains near background levels ($<10$ $\mu$g m$^{-3}$) except for a few short spikes, while NO\textsubscript{2} is likewise low save for few dramatic spikes above $100$. 
From these plots for each city, a clear trend emerges: PM\textsubscript{2.5} and PM\textsubscript{10} levels show a strong positive correlation, indicating that they tend to rise and fall together. In contrast, PM\textsubscript{2.5} levels (also the other pollutants) appear to be inversely related to relative humidity (RH), suggesting an anti-correlation where higher humidity is generally associated with lower PM\textsubscript{2.5} concentrations.

\subsection{Nonlinearity in Correlation}
To complement the time-series diagnostics, we next explore the instantaneous relationships between each air-quality indicator (PM\textsubscript{2.5}, PM\textsubscript{10}, NO\textsubscript{2}, SO\textsubscript{2}) and the two meteorological drivers (RH, AT). For every pollutant–weather variable pair we pool the full five-year daily dataset and fit a non-parametric regression smoothing with a kernel regression framework~\cite{non_param1}.
In statistics, kernel regression is a non-parametric method used to detect the type of relationship present between random variables $X$ and $Y$. The regression takes the form,
\begin{equation}
    E(Y|X)=m(X)
\end{equation}
where $E(Y|X)$ denotes the expectation value of $Y$, conditioned on $X$, and $m$ is an unknown function. A common estimator for this function is the Nadaraya–Watson estimator,
\begin{equation}
    \hat{m}(x)=\frac{\sum_{i=1}^NK_h(x-X_i)Y_i}{\sum_{i=1}^NK_h(x-X_i)},
\end{equation}
with $Y_i$ as the pollutant concentration, $X_i$ the weather variable, and $K_h(z)$ is a Gaussian kernel with bandwidth $h$ selected via leave-one-out cross-validation~\cite{non_param,non_param2}. This approach places more weight on observations whose meteorological conditions are close to the point of interest, without imposing a parametric shape.

\begin{figure*}[tbh]
    \centering    
\includegraphics[width=0.92\linewidth]{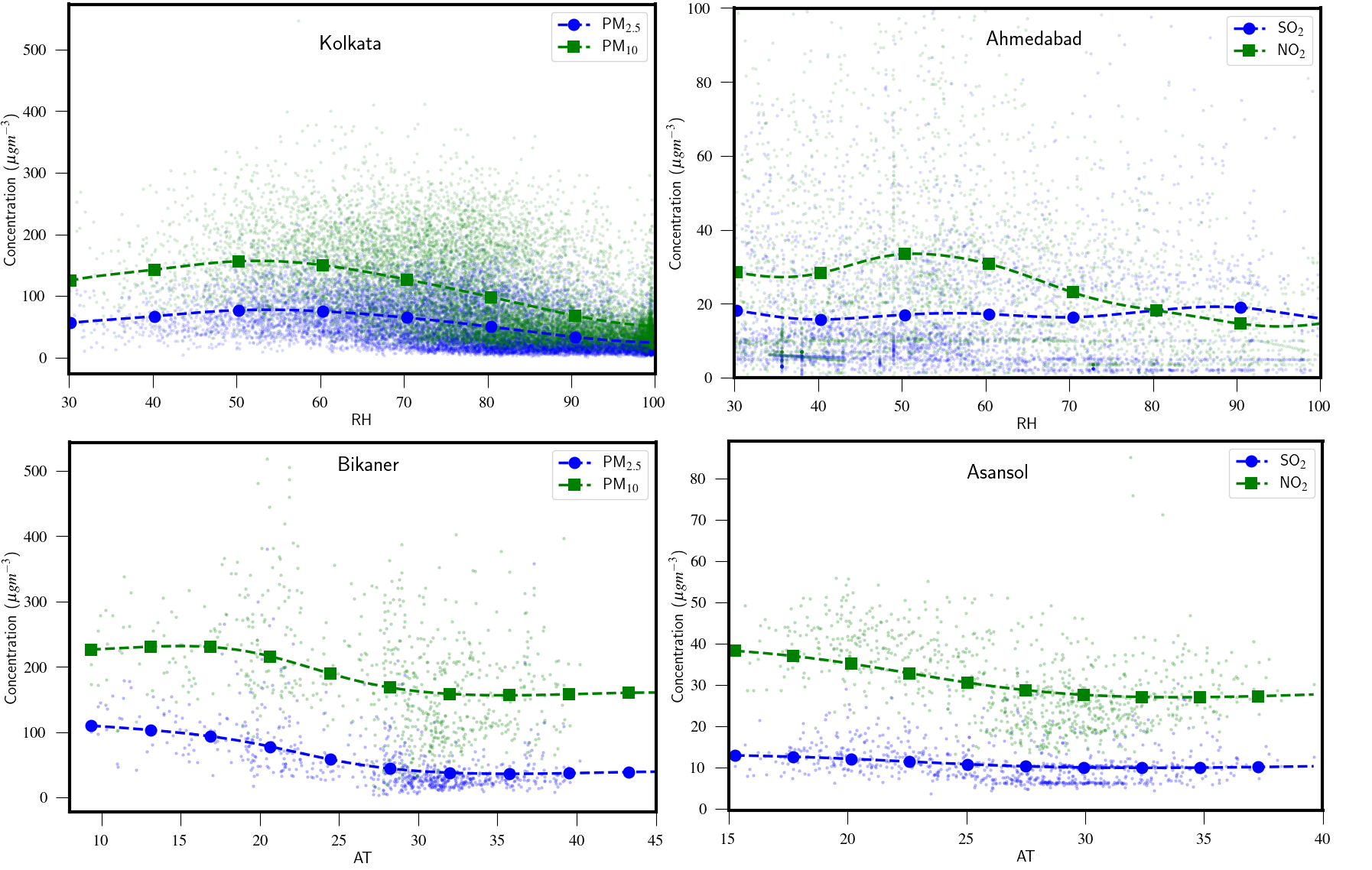} 
    \caption{Kernel regression plots illustrating relationships between pollutants and meteorological variables in selected  cities.}
    \label{fig:non_params}
\end{figure*}

Figure~\ref{fig:non_params} show the kernel regression plots for different pairs of pollutant-meteorological parameters for a few representative cities. These plots
demonstrate clear nonlinear relationships between the pollutants and meteorological variables, indicating that traditional linear methods may not adequately assess  the complexity of these interactions.

In the following section, we characterize this non-trivial relationship between the pollutants and the weather variables using several different measures.

\section{Correlation of Pollutant and Weather variables}\label{sec:corr}

In this section we focus on the core aspect of our study evaluating the strength of interaction between pollutant and meteorological variable pairs across different cities. This includes assessing both linear and non-linear associations, thereby providing insight into how these relationships vary spatially. Such an analysis enables a classification of cities based on the degree of correlation among the variable pairs.

To quantify these associations, we compute three complementary measures, namely, the Pearson correlation coefficient, mutual information, and relative conditional entropy between each pollutant-weather variable pair.  While the Pearson coefficient estimates the linear dependencies, the other two entropy based measures are more general, and are able to characterize non-linear and asymmetric dependencies.


To synthesize these diverse metrics into a unified measure of correlation, we introduce a composite correlation index. 
This index is calculated as the arithmetic mean, (i.e., unweighted average) of the three normalized measures; Pearson correlation, mutual information and relative conditional entropy, thereby integrating multiple 
metrics into a single interpretable metric. To the best of our knowledge, this represents the first attempt to derive such a composite correlation index in a straightforward yet comprehensive manner.

\subsection{Pearson Correlation Coefficient}
We begin by estimating the linear correlation between each pollutant–meteorological variable pair for all the cities.  The simplest quantitative estimate of the linear interdependency of two observables is the Pearson correlation coefficient (PCC)~\cite{gun2008fundamentals}. The PCC between two time series $X=\{x_{1},x_{2},..., x_{N}\}$ and $Y=\{y_{1},y_{2},...,y_{N}\}$ is defined as,
\begin{align}
  r_{X,Y}=\frac{\sum_{i=1}^{N}(x_i-\hat{x})(y_i-\hat{y})}{\sqrt{\sum_{i=1}^{N}(x_i-\hat{x})^{2}}\,\sqrt{\sum_{i=1}^{N}(y_i-\hat{y})^{2}}},  
\end{align}
where, $\hat{x}=\frac {1}{N} \sum_{i=1}^{N}x_i$ and $\hat{y}=\frac {1}{N} \sum_{i=1}^{N}y_i$ are the means of the respective time-series. 
Here $n$ denotes the length of the time-series, which, for our case varies from station to station. Clearly, PCC is symmetric under the exchange of $X$ and  $Y$ and its values is bounded by $-1 \le r_{X,Y} \le 1$. While a perfect positive (negative) linear correlation between $X$ and $Y$ is indicated by $r_{X,Y}=1$ ($r_{X,Y}=-1$), $r_{X,Y}=0$ indicates absence of linear correlation.  We compute the $r_{X,Y}$ values between the eight pollutant--weather variable pairs, namely, PM\textsubscript{2.5}-RH, PM\textsubscript{2.5}-AT, PM\textsubscript{10}-RH etc., for the selected cities. This is summarized in Table~\ref{tab:pearson}.

It appears that $r_{\text{RH},\text{PM}_{2.5}}<0$ for most of the cities, which indicates that relative humidity and PM\textsubscript{2.5} affect each other negatively, at least to linear order. A similar trend is observed for PM\textsubscript{10}, with RH showing a negative correlation with it in most locations, reinforcing the idea that moisture in the air can suppress particulate matter levels. Ambient temperature (AT), too, tends to correlate negatively with PM\textsubscript{2.5} and PM\textsubscript{10} across many cities, especially in urban centers like Kolkata, Delhi, Mumbai and Asansol, where cooler temperatures may coincide with poor dispersion conditions or increased local emissions, leading to pollutant accumulation. For gaseous pollutants like SO\textsubscript{2} and NO\textsubscript{2}, the relationships are more varied. While several cities show weak or inconsistent correlations, others (e.g., Kolkata, Asansol, Bhubaneswar, and Bikaner) display significant negative correlations. Overall, both RH and AT have a measurable influence on air pollutant concentrations, particularly for particulate matter, and that these meteorological variables should be considered when analyzing or forecasting urban air quality.

\begin{table*}[th]
\centering
\small
    \begin{tabular}{|c||c|c|c|c|c|c|c|c|}
\hline
\textbf{City} & $r_{\text{\tiny RH},\text{\tiny PM}_{2.5}}$ & $r_{\text{\tiny AT},\text{\tiny PM}_{2.5}}$ & $r_{\text{\tiny RH},\text{\tiny PM}_{10}}$ & $r_{\text{\tiny AT},\text{\tiny PM}_{10}}$ & $r_{\text{\tiny RH},\text{\tiny SO}_{2}}$ & $r_{\text{\tiny AT},\text{\tiny SO}_{2}}$ & $r_{\text{\tiny RH},\text{\tiny NO}_{2}}$ & $r_{\text{\tiny AT},\text{\tiny NO}_{2}}$ \\
\hline
    AMD& -0.27& -0.17& -0.27& -0.12& \cellcolor{blue!25}0.04& \cellcolor{blue!25}0.07&-0.1& \cellcolor{blue!25}0.02
\\
    \hline
    BLR& -0.14& \cellcolor{blue!25}0.05& -0.27& \cellcolor{blue!25}0.12&\cellcolor{blue!25} 0.09& \cellcolor{blue!25}0.04&-0.19& -0.04
\\
    \hline
    CHN& \cellcolor{blue!25}0.11&-0.29&-0.08&-0.21& \cellcolor{blue!25}0.06& -0.14& \cellcolor{blue!25}0.04&-0.08
\\
    \hline
    DEL& -0.08&-0.37& \cellcolor{blue!25}0.15& -0.52&-0.13& -0.13& -0.06& -0.17
\\
    \hline
    HYD& -0.34& -0.14& -0.44& -0.06& \cellcolor{blue!25}0.11&\cellcolor{blue!25} 0.20& -0.24&-0.22
\\
    \hline
    KOL& -0.42& -0.50&-0.47& -0.49& -0.30& -0.09& -0.24& -0.34
\\
    \hline
    MUM&  -0.52& -0.20& -0.59& -0.06&  \cellcolor{blue!25}0.01& \cellcolor{blue!25} 0.05&  -0.45&  -0.10
\\
    \hline
    AGT& \cellcolor{blue!25} 0.10& -0.24&-0.02& -0.28& -0.16& -0.30& \cellcolor{blue!25} 0.03&  -0.15
\\
    \hline
    ASN&  -0.35& -0.49& -0.39& -0.53& -0.51& -0.28& -0.43& -0.49
\\
    \hline
    BDI& \cellcolor{blue!25}0.09& -0.58& -0.13& -0.37& -0.04& -0.41& \cellcolor{blue!25}0.10&-0.62
\\
    \hline
    BGP& -0.03& -0.63& -0.24& -0.39&  \cellcolor{blue!25}0.13&  \cellcolor{blue!25}0.06& -0.16& -0.25
\\
    \hline
 BPL& -0.19&-0.54& -0.38& -0.38& -0.15& \cellcolor{blue!25}0.05& -0.44&-0.38
\\ \hline
    BSR& -0.28&-0.62& -0.32& -0.54& -0.28&-0.12& -0.23& -0.71
\\
    \hline
    BKN& \cellcolor{blue!25}0.01& -0.54&-0.41& -0.36& -0.27&-0.34& -0.32&-0.35
\\
    \hline
    GWH&-0.08&-0.48& -0.12& -0.42&  \cellcolor{blue!25}0.31& \cellcolor{blue!25} 0.12& \cellcolor{blue!25} 0.02& -0.30
\\
    \hline
    IND& -0.20& -0.30& -0.51& -0.11& -0.49&\cellcolor{blue!25} 0.13& -0.54& -0.18
\\
    \hline
    KOP& -0.49&  \cellcolor{blue!25}0.02& -0.62&  \cellcolor{blue!25}0.07& -0.21&  \cellcolor{blue!25}0.03& -0.58&  \cellcolor{blue!25}0.05
\\
    \hline
    LKO&-0.12&-0.39& -0.30& -0.28&  \cellcolor{blue!25}0.08& \cellcolor{blue!25} 0.10& -0.20& -0.18
\\
    \hline
    MAN& \cellcolor{blue!25} 0.09& \cellcolor{blue!25} 0.001& \cellcolor{blue!25} 0.14& -0.01&  -0.37&-0.09&-0.47& \cellcolor{blue!25} 0.30
\\
    \hline
    PDY&-0.02& -0.30& -0.30& -0.09& -0.25& \cellcolor{blue!25} 0.25& -0.01&-0.44
\\
    \hline
    RPR&-0.35&-0.23& -0.38& -0.18& -0.32& \cellcolor{blue!25} 0.34&  -0.52& -0.17
\\
    \hline
    SXR& -0.10& \cellcolor{blue!25}0.01&-0.02& -0.07&-0.09& -0.10& -0.03& -0.25
\\
    \hline
    TVM& -0.07&\cellcolor{blue!25} 0.33& -0.04& \cellcolor{blue!25}0.45& -0.42&-0.6& \cellcolor{blue!25}0.12& \cellcolor{blue!25}0.16
\\
    \hline
 VTZ&-0.18&-0.43& -0.16& -0.21& \cellcolor{blue!25}0.16& -0.03&-0.15&-0.1
\\ \hline
    \end{tabular}
    \caption{Pearson Correlation Coefficients of pollutants and weather variables across cities with maximum $7\%$ margin of error. Blue colored cells highlight positive correlation.}
    \label{tab:pearson}
    \end{table*}

\subsection{Mutual Information}
To further explore the dependence structure between air pollutants and weather variables, we compute the mutual information~\cite{Steu_2002,Kras_2004}. 
MI quantifies the reduction in uncertainty of one variable given knowledge of the other, characterizing both linear and non-linear dependencies. The mutual information  between air pollutants ($X$) and meteorological factors ($Y$) is defined as 
 \begin{equation}
I_{X,Y}=S_X+S_Y-S_{X,Y}\,\,,
\end{equation}
where $S_X$ and  $S_Y$ represent the self-entropy (Shannon Entropy) of variables $X$ and $Y$, respectively, and $S_{X,Y}$ denotes their joint entropy.


Let $p(x)$ denote the probability density function (PDF) of the stationary time-series $X = \{x_t; t=1, \cdots N\}$. The corresponding Shannon entropy, which quantifies the average uncertainty or information content in a PDF, is defined as, 
\begin{equation}
    S_X = -\int dx\,p(x)\,\ln[p(x)]. \label{eq:Sx}
\end{equation} 
Higher entropy values indicate greater  uncertainty  or disorder in the  system. The joint entropy  $S_{X,Y}$, which quantifies the combined uncertainty of both variables, is defined as,
\begin{equation}
   S_{X,Y}=-\int dx \int dy\,p(x,y)\,\ln[p(x,y)], \label{eq:Sxy}
\end{equation} 
where $p(x,y)$ denotes the joint probability density of $X$ and $Y$.

We construct the PDF of each observable (all pollutants and weather variables) from the aggregated data from all the stations in each city. The joint distributions of each pollutant-weather variable pair is also constructed from the same data. The self-entropies of each observable and the mutual information of the pollutant-weather variables pairs are computed from these distributions.  


Table~\ref{tab:Entropy} summarizes the self-entropy values of all the pollutant and weather variables across all cities, which indicate their individual uncertainty levels. It is apparent from this Table that particulate matter shows the greatest variability, with $S_{\text{PM}_{10}}$ typically highest (lying in the range $ 4.4-5.8$ with the maximum in Bhagalpur) and $S_{\text{PM}_{2.5}}$ close behind (in the range between $3.9-6.2$; exceptional peak in Delhi). Gaseous pollutants are lower and more site-specific,$S_{\text{NO}_2}$ sits mostly between $3.0-4.5$ (highest in Indore and Delhi), while $S_{\text{SO}_2}$ is the smallest among pollutants (in the range $1.8-4.0$, notably low in Kolhapur). Among the meteorological parameters, relative humidity (RH) shows moderate spread (lying in the range $3.1-4.4$; higher in Ahmedabad, Bhopal and Raipur), whereas ambient temperature (AT) has the narrow spreading (ranging from $2.1-3.3$) with one outlier in Mangaluru ($0.546$), indicating very limited temperature variability there (or possible data issues). In short, across cities the uncertainty ranking is roughly $S_{\text{PM}_{10}}\geq S_{\text{PM}_{2.5}}>S_{\text{NO}_{2}}\sim S_{\text{RH}}>S_{\text{SO}_{2}}>S_{\text{AT}}$, highlighting that particulates fluctuate most from day to day, while temperature is comparatively stable. 

\par Mutual Information $I_{X,Y}$ is then calculated  for all  variable pairs, across all  cities  to quantify  the total amount of  shared information between pollutant and meteorological variables. The evaluated values of $I_{X,Y}$ between pollutants and weather variables across multiple cities are presented in Table \ref{tab:MI}.

This table summarizes mutual information between each pollutant and the two meteorological drivers (RH, AT) for all cities. Broadly, particulates show the strongest weather links, often tighter with humidity than temperature: e.g., $I_{\text{PM},\text{RH}}$ exceeds $I_{\text{PM},\text{AT}}$ in Ahmedabad, Bengaluru, Hyderabad, Mumbai, Kolhapur, etc., while a few places (Chennai, Kolkata) show comparable or higher $I_{\text{PM},\text{AT}}$. Several cities stand out with consistently high $I$ across many pairs notably Asansol (lying in the range $0.47-0.60$ for most pairs), Agartala, Baddi, Bhubaneswar, Bikaner, and Kolhapur for $I_{\text{PM}_{10},\text{RH}}$ ($0.617$). The largest value of $I$ in the table is $I_{\text{NO}_{2},\text{AT}}$ in Bhubaneswar ($0.755$), indicating strong temperature control on NO\textsubscript{2} there; $I_{\text{PM}_{10},\text{RH}}$ in Bikaner ($0.703$) and PM pairs in Asansol ($0.60$) are also prominent. For SO\textsubscript{2}, the strongest dependencies tend to be with AT in certain climates e.g., Mangaluru ($0.599$), Thiruvananthapuram ($0.416$), Srinagar ($0.293$) suggesting thermally driven chemistry or sources, whereas other cities show modest SO\textsubscript{2}–weather coupling. Coastal metros like Mumbai display higher $I_{\text{PM}–\text{RH}}$ ($0.268-0.305$) but weaker $I_{\text{PM}–\text{AT}}$, consistent with moist boundary-layer effects. Overall, $I$ values are city-specific, but the common pattern is $I_{\text{PM} \leftrightarrow \text{RH}} > I_{\text{PM} \leftrightarrow \text{AT}}$, while NO\textsubscript{2} often tracks AT, and SO\textsubscript{2} shows mixed but sometimes temperature-led behavior.
\begin{table*}[th]
    \centering
    \small
    \begin{tabular}{|c||c|c|c|c|c|c|c|c|}
\hline
{\textbf{City}} & $I_{\text{\tiny PM}_{2.5},\text{\tiny RH}}$ & $I_{\text{\tiny PM}_{2.5},\text{\tiny AT}}$ & $I_{\text{\tiny PM}_{10},\text{\tiny RH}}$ & $I_{\text{\tiny PM}_{10},\text{\tiny AT}}$ & $I_{\text{\tiny SO}_{2},\text{\tiny RH}}$ & $I_{\text{\tiny SO}_{2},\text{\tiny AT}}$ & $I_{\text{\tiny NO}_{2},\text{\tiny RH}}$ & $I_{\text{\tiny NO}_{2},\text{\tiny AT}}$ \\
\hline
AMD&0.180 &0.144 &0.196 &0.151 &0.153 & 0.106& 0.162& 0.136 \\\hline
 BLR& \cellcolor{red!25}0.073& \cellcolor{red!25}0.023& 0.117& \cellcolor{red!25}0.059& 0.064& 0.030& 0.097&\cellcolor{red!25}0.047
\\
    \hline
    CHN&  0.076& 0.110& \cellcolor{red!25}0.054&  0.112&  0.082& 0.088& 0.069& 0.091
\\
    \hline
    DEL& 0.105& 0.240& 0.092& 0.139& 0.060& 0.022& \cellcolor{red!25}0.017& 0.078
\\
    \hline
    HYD& 0.204& 0.073& 0.234& 0.076& 0.082& 0.127& 0.128& 0.111
\\
    \hline
    KOL& 0.212& 0.265& 0.221& 0.269& 0.043& \cellcolor{red!25}0.018& 0.082& 0.104
\\
    \hline
    MUM& 0.268& 0.071& 0.305& 0.061& \cellcolor{red!25}0.031& 0.080& 0.166& 0.067
\\
    \hline
    AGT& 0.400& 0.515& 0.360& 0.554& 0.272& 0.406& 0.389& 0.502
\\
   \hline
    ASN& \cellcolor{blue!25}0.597&0.602 & 0.599&0.570 &\cellcolor{blue!25}0.555 &0.534&0.475&0.536 \\
    \hline
    BDI&0.329 &0.516 &0.347 & 0.439&0.456 &0.499 &0.357 &0.531 \\
    \hline
    BGP&0.280 &0.414 & 0.279&0.296 & 0.120& 0.156&0.215 & 0.289\\
    \hline
    BPL& 0.333& 0.380& 0.381& 0.332&0.223& 0.256&0.403&0.361
\\\hline
BSR& 0.496&\cellcolor{blue!25}0.726&0.488&\cellcolor{blue!25}0.653&0.274&0.355&0.504&\cellcolor{blue!25}0.755
\\
        \hline
    BKN&0.521 & 0.559&\cellcolor{blue!25}0.703 &0.535&0.333&0.350&0.530&0.499\\
        \hline
    GWH& 0.348& 0.470& 0.385& 0.469& 0.474&0.439&0.242&0.325
\\
        \hline
    IND&0.244&0.241&0.451&0.304&0.254&0.159&0.406&0.308
\\
        \hline
    KOP&0.471 &0.196 &0.617&0.310&0.240&0.086&\cellcolor{blue!25}0.553&0.246\\
        \hline
    LKO&0.145 &0.198&0.180&0.160&0.203&0.184&0.118&0.103\\
        \hline
    MAN&0.311 & 0.230& 0.339& 0.293&0.512& \cellcolor{blue!25}0.599&0.297&0.309\\
        \hline
    PDY&0.208 &0.229 & 0.301&0.245 &0.260&0.208 &0.208&0.306\\
        \hline
    RPR& 0.253&0.163&0.301&0.220&0.184&0.195&0.317&0.206
\\
        \hline
    SXR&0.222&0.272&0.247&0.298&0.176&0.293&0.312&0.450\\
\hline
    TVM&0.183& 0.253&0.144&0.253&0.263&0.416&0.110&0.158
\\
        \hline
 VTZ
& 0.196& 0.292& 0.209& 0.267& 0.162& 0.166& 0.162&0.216
\\ \hline 
    \end{tabular}
    \caption{Mutual Information values for eight pollutant–meteorological variable pairs across cities. For each pair, the maximum and minimum MI values are highlighted in blue and red, respectively.}
    \label{tab:MI}
\end{table*}

\subsection{Relative Conditional Entropy}
To further estimate the strength, complexity and conditional structure of interactions between air pollutants ($X$) and meteorological factors ($Y$), (both linear and non-linear), we compute relative conditional entropy.
The conditional entropy for a pollutant ($X$), given a weather variable ($Y$) is defined as,
\begin{equation}\label{eq:7}
   {\cal H}_{X;Y} = S_{X,Y} - S_{Y},
\end{equation}
where $S_{X,Y}$ and $S_Y$ denote the joint entropy of $X$ and $Y$ [see Eq.~\eqref{eq:Sxy}] and self-entropy of $Y$ [see Eq.~\eqref{eq:Sx}], respectively.
To characterize the fraction of uncertainty remaining in $X$ after conditioning on $Y$, it is useful to define a relative conditional entropy as,
\bea
{\cal H}^R_{X;Y} = {\cal H}_{X;Y}/ S_X.
\eea 
It can be easily shown that $ 0 \le {\cal H}^R_{X;Y} \le 1$. Smaller value of ${\cal H}^R_{X;Y}$ indicates that $X$ has a strong dependence on $Y$, while 
${\cal H}^R_{X;Y}=1$ indicates complete independence of the two variables. 

The values of ${\cal H}$ and ${\cal H}^R$ computed from the aggregated data across all the chosen cities are tabulated in Tables~\ref{tab:Cond_Entropy} and \ref{tab:relative} respectively. Table~\ref{tab:Cond_Entropy} lists the conditional entropy ${\cal H}$ i.e., the uncertainty left in each pollutant after conditioning on RH or AT. 
Blue-highlighted maxima (e.g., Delhi for PM\textsubscript{2.5}, PM\textsubscript{10}, NO\textsubscript{2}) indicate weak control, while red minima (e.g., Guwahati for NO\textsubscript{2}, Kolhapur for SO\textsubscript{2}, Puducherry and Thiruvananthapuram for PM) show cities where meteorology removes the most uncertainty; PM\textsubscript{10} tends to be slightly less meteorology-driven than PM\textsubscript{2.5}. The relative conditional entropy ${\cal H}^{R}$ (i.e, fraction of uncertainty remaining) in Table~\ref{tab:relative} echoes the same pattern: values near $1$ (blue; e.g., Delhi ${\cal H}^R_{\text{PM}_{10};\text{RH}/\text{AT}}$, several Bengaluru pairs, Mumbai ${\cal H}^R_{\text{SO}_{2};\text{RH}}$) mean meteorology explains little, whereas small red values (e.g., Kolhapur ${\cal H}^R_{\text{SO}_{2};\text{RH}}$, Guwahati ${\cal H}^R_{\text{NO}_{2};\text{RH}/\text{AT}}$, BSR ${\cal H}^R_{\text{NO}_{2};\text{AT}}$ mean RH/AT account for a substantial share of pollutant variability. Together, these two tables say: where ${\cal H}$ is low and ${\cal H}^R$ is small, meteorology (RH/AT) is highly informative about pollutant levels; where they are high or near $1$, pollutants vary largely independent of RH/AT.

\begin{table*}[th]
\centering
 \small
    \begin{tabular}{|c||c|c|c|c|c|c|c|c|}
\hline
{\textbf{City}} & ${\cal H}^R_{\text{\tiny PM}_{2.5};\text{\tiny RH}}$ & ${\cal H}^R_{\text{\tiny PM}_{2.5};\text{\tiny AT}}$ & ${\cal H}^R_{\text{\tiny PM}_{10};\text{\tiny RH}}$ & ${\cal H}^R_{\text{\tiny PM}_{10};\text{\tiny AT}}$ & ${\cal H}^R_{\text{\tiny SO}_{2};\text{\tiny RH}}$ & ${\cal H}^R_{\text{\tiny SO}_{2};\text{\tiny AT}}$ & ${\cal H}^R_{\text{\tiny NO}_{2};\text{\tiny RH}}$ & ${\cal H}^R_{\text{\tiny NO}_{2};\text{\tiny AT}}$\\
\hline
AMD
& 0.963& 0.987& 0.967& 0.990& 0.890& 0.937& 0.891& 0.936
\\
    \hline
    BLR
& \cellcolor{blue!25}0.989&\cellcolor{blue!25} 1.0& 0.976& 0.992& 0.973& \cellcolor{blue!25}0.998& 0.974& 0.991
\\
    \hline
    CHN
& 0.952& 0.959& 0.971& 0.972& 0.983& 0.985& 0.953& 0.969
\\
    \hline
    DEL
& 0.887& 0.866& \cellcolor{blue!25}1.0& \cellcolor{blue!25}1.0& 0.839& 0.850& \cellcolor{blue!25}1.0& \cellcolor{blue!25}0.998
\\
    \hline
    HYD
& 0.955& 0.983& 0.955& 0.985& 0.978& 0.967& 0.972& 0.977
\\
    \hline
    KOL
& 0.972& 0.914& 0.971& 0.922& 0.995& 0.943& 0.993& 0.942
\\
    \hline
    MUM
& 0.958& 0.984& 0.956& 0.990&\cellcolor{blue!25}1.0& 0.988&0.975&0.986
\\
   \hline
    AGT
& 0.916& 0.921& 0.942& 0.935& 0.944&0.939&0.946&0.941
\\
        \hline
    ASN
& \cellcolor{red!25}0.855& 0.930& \cellcolor{red!25}0.858& 0.933&0.839& 0.951&0.848&0.916
\\
        \hline
    BDI
& 0.902& 0.925& 0.936& 0.972&0.914& 0.952&0.943&0.928
\\
        \hline
        BGP
& 0.934& 0.910& 0.961& 0.962&0.968& 0.972&0.905&0.914
\\
        \hline
 BPL
& 0.933& 0.925& 0.944& 0.961& 0.974& 0.988& 0.940&0.965
\\ \hline
BSR
& 0.918& \cellcolor{red!25}0.861& 0.917& \cellcolor{red!25}0.889&0.913& 0.892&0.847&0.766
\\
        \hline
    BKN
& 0.896& 0.894& 0.942& 0.972&0.899& 0.924&0.941&0.961
\\
        \hline
    GWH
& 0.923& 0.902& 0.933& 0.921&0.853& 0.872&\cellcolor{red!25}0.725&\cellcolor{red!25}0.645
\\
        \hline
    IND
& 0.911& 0.956& 0.913& 0.976&0.896& 0.991&0.901&0.978
\\
        \hline
    KOP
& 0.879& 0.971& 0.891& 0.978&\cellcolor{red!25}0.577& 0.828&0.873&0.973
\\
        \hline
    LKO
& 0.955& 0.967& 0.949& 0.975&0.958& 0.987&0.913&0.949
\\
        \hline
    MAN
& 0.972& 0.990& 0.979& 0.986&0.938& 0.959&0.885&0.936
\\
        \hline
    PDY
& 0.964& 0.954& 0.975& 0.984&0.986& 0.989&0.986&0.936
\\
        \hline
    RPR
& 0.934& 0.970& 0.946& 0.981&0.910& 0.939&0.887&0.943
\\
        \hline
    SXR
& 0.985& 0.969& 0.981& 0.965&0.956& 0.950&0.948&0.923
\\
        \hline
    TVM
& 0.956& 0.930& 0.968& 0.939&0.878& \cellcolor{red!25}0.814&0.961&0.943
\\
        \hline
 VTZ
& 0.944& 0.919& 0.954& 0.940& 0.924& 0.918& 0.949&0.931
\\ \hline        
    \end{tabular}
    
    \caption{Relative conditional entropy values across cities with maximum (blue) and minimum (red) ${\cal H}^R$ values highlighted for each pair.}
    \label{tab:relative}
\end{table*}


\subsection{Composite Correlation Index}
The different metrics discussed so far, namely, Pearson correlation, relative conditional entropy and MI,  complement each other in characterizing the interdependence of the pollutants and weather variables. To gain a complete understanding of this interdependence, it is useful to define a single metric which combines the three distinct metrics.  Such a multi-metric approach ensures a comprehensive assessment by evaluating linear and nonlinear associations, as well as potentially asymmetric dependencies.


To develop a unified measure of the correlations, we first normalize all the metrics  to a common scale ranging from $0$ to $1$ using Min-Max normalization~\cite{Panda_2015}.  For each variable pair $(X,Y)$, the normalized version of the  metric $M_{XY}$ (where $M_{XY}$ indicates either of $r_{X,Y}$, $I_{X,Y}$ and ${\cal H}^R_{X;Y}$) is defined as, 
\begin{equation}
    \tilde{M}_{XY} = \frac{M_{XY} - \min(\vec M_{XY})}{\max(\vec M_{XY})-\min(\vec M_{XY})} ,
\end{equation}
where  the  $j$-th entry of the vector $\vec M = \{ M_j; j=1, \cdots 24 \}$ denotes the value of the relevant metric for the $j$-th city. The normalized metrics are then aggregated via arithmetic mean to produce a composite correlation index (CCI) for each variable pair as,
\begin{equation}
   C_{X,Y} = \frac{1}{3}[\tilde{r}_{X,Y}+\tilde{I}_{X,Y}+(1-\tilde{\cal H}^R_{X,Y})] 
\end{equation}
where, $\Tilde{r}$, $\Tilde{I}$ and $\Tilde{\cal H}^R$  are the normalized values of the respective metrics\footnote{Note that, larger $\Tilde{\cal H}^R$ indicates weaker interaction and hence taking $(1-\Tilde{\cal H}^R)$ flips its direction so that all three terms are on the same scale}.  
Note that, $0 \le C_{X,Y} \le 1$, with higher values indicating higher overall correlation between the variable pair $X,Y$ in a city. Table~\ref{tab:comp} summarizes the CCI values for all the variable pairs in all the cities. \\



\noindent \textbf{Classification of cities based on CCI values:}
To further interpret the interdependency landscape, we focus on four pollutant–meteorological variable pairs from the broader set --- PM\textsubscript{2.5}–RH, PM\textsubscript{2.5}–AT, NO\textsubscript{2}–RH and NO\textsubscript{2}–AT. These are analyzed across all the twenty four locations. The choice of these representative pollutants is guided by their strong relevance to urban air quality and public health as well as  their differing physical characteristics.  
The CCI values of these variable pairs is now  used to classify the cities into  distinct  groups.

\begin{table}[th]
    \centering
    \begin{tabular}{|p{1.0 cm}|p{1.0 cm}|p{1.0 cm}|p{1.1 cm}|p{1.0 cm}|}
    \hline 
     $C_{X,Y}$  & Very High   & High & Moderate & Low   \\
        \hline 
        \hline
      \vspace{0.8 cm} PM\textsubscript{2.5} - RH   & BKN, BDI, ASN, AGT &
      DEL, BGP, BPL, BSR, GWH, IND, KOP, MAN & CHN, LKO, PDY, RPR, SXR, TVM, VTZ  & AMD, BLR, HYD, KOL, MUM \\ 
      \hline 
     \vspace{0.8 cm}  PM\textsubscript{2.5} - AT   & BSR & DEL, AGT, ASN, BKN, GWH, TVM  & KOL, BDI, BGP, BPL, IND, KOP, MAN, PDY, SXR, VTZ & AMD, BLR, CHN, HYD, MUM, LKO, RPR\\ 
      \hline 
     \vspace{0.8 cm}  NO\textsubscript{2}- RH & BSR, GWH & AGT, ASN, BDI,
BKN, KOP, SXR & AMD, CHN, BGP, BPL, IND, LKO, MAN, PDY, RPR, TVM, VTZ & BLR, DEL, HYD, KOL, MUM \\
      \hline 
     \vspace{0.8 cm}  NO\textsubscript{2}- AT & -- & BSR, GWH, MAN  & AMD, AGT, ASN, BDI, BGP, BKN, IND, KOP, SXR, TVM, VTZ & BLR, CHN, DEL, HYD, KOL, MUM, BPL, LKO, PDY, RPR \\
      \hline
    \end{tabular}
    \caption{Clustering of cities based on the values of the composite correlation index for four variable pairs}.\label{tab:couplings}
    \label{tab:placeholder}
\end{table}

To derive objective and data-driven thresholds for categorizing the CCI values, we apply k-means clustering~\cite{k-means} algorithm  to the set of $C$ values for the relevant variable pairs across 24 cities. This unsupervised approach allows us to partition the data into four distinct regimes based on natural groupings within the data. Compared to arbitrary thresholds, this method ensures that the classification reflects inherent variability and distribution of correlation index, providing  statistically grounded framework for interpreting spatial differences in pollutant–meteorology interactions.
Table~\ref{tab:couplings} shows the classification of cities on the basis of $C$ values:\\
\begin{itemize}
    \item Very high correlation: $C > 0.66$
    \item High correlation: $0.47 < C \leq 0.65$
    \item Moderate correlation: $0.315 < C \leq 0.46$
    \item Low correlation: $C \leq 0.305$
\end{itemize}
for each pollutant-weather variable pair.

\begin{figure}[tbh]
\includegraphics[width=0.8\linewidth]{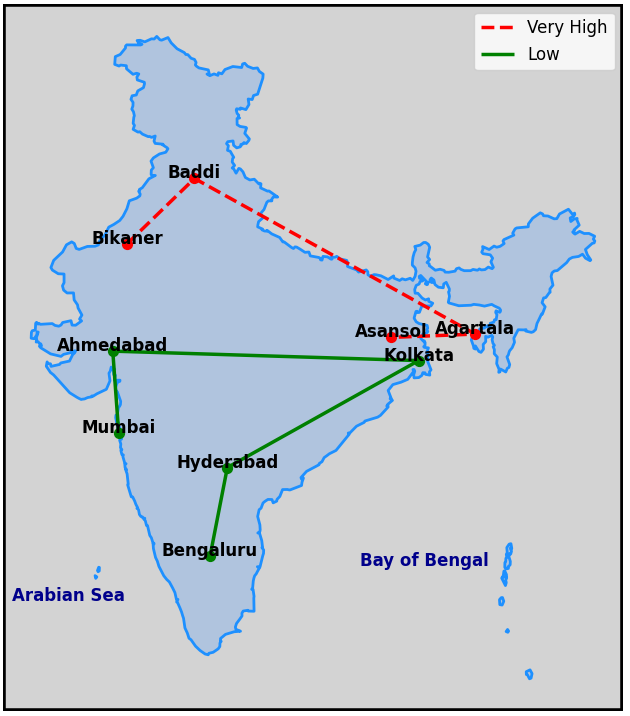}
\includegraphics[width=0.8\linewidth]{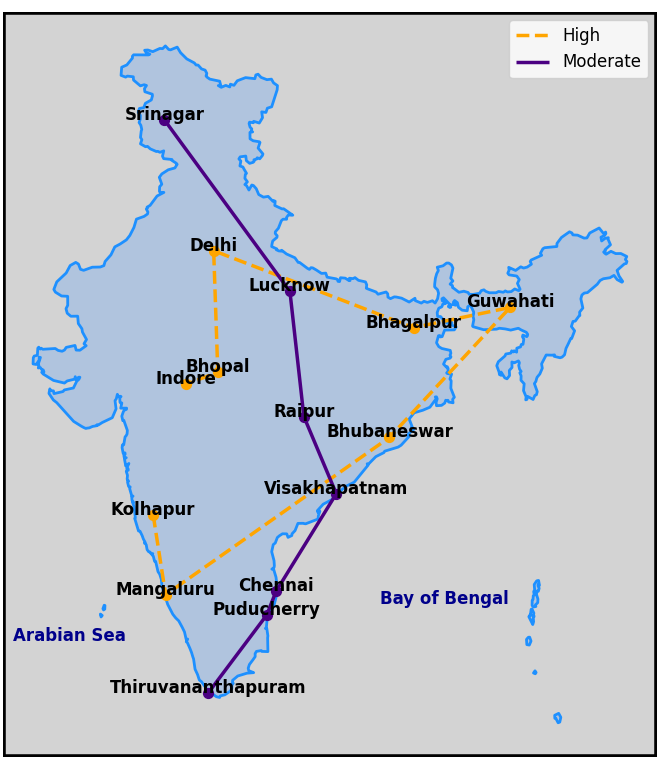}
\caption{Geographic visualization of city clusters based on PM\textsubscript{2.5}–RH correlation. Cities in the same group are connected by lines. The top panel shows the cities corresponding to very high (dashed red) and low (solid green)  CCI values while the  bottom panel shows cities with high (dashed orange)  and moderate (solid violet) values of CCI }.
\label{fig:directed map}
\end{figure}

Figure~\ref{fig:directed map} displays the geographic visualization of the cities within each cluster, based on the PM\textsubscript{2.5}–RH correlation identified in Table~\ref{tab:couplings}. This highlights the spatial coherence within each cluster.

\section{Temporal Characteristics of PM\textsubscript{2.5}-RH Interdependence}\label{sec:temp}

The preceding analyses reveal that, among all pollutant–meteorological pairings, PM\textsubscript{2.5} exhibits the most pronounced statistical association with RH. Motivated by this observation, the present section focuses exclusively on unraveling the temporal dynamics underlying the PM\textsubscript{2.5}–RH relationship. To investigate the directionality and delayed dependencies we employ two complementary techniques in the time domain:
Transfer Entropy (TE), which quantifies the amount of information that past values of RH convey about the future evolution of PM\textsubscript{2.5} (and vice versa) beyond what is already contained in each series’ own history, thereby indicating possible causal influence, and Time-Delayed Mutual Information (TDMI), which measures the overall (linear and non-linear) mutual dependence between the two variables as a function of temporal lag. 
Together, these metrics furnish a detailed portrait of how RH modulates PM\textsubscript{2.5} concentrations over time helping to disentangle mere correlation from genuine predictive or causal relationships.


\subsection{Causal Interaction: Transfer Entropy} 
We utilize transfer entropy~\cite{TE_2000,TE_2011}, a non-parametric  information-theoretic measure to explore the directional dependencies and potential causal interactions between air pollutants and meteorological parameters. The transfer entropy from a process $Y$ to another process $X$ is defined as,
\begin{align}
{\cal T}_{Y\xrightarrow{}X}&={\cal H}_{X_{t};X_{t-1}}-{\cal H}_{X_{t};X_{t-1},Y_{t-1}}, \label{eq:Txy}
\end{align}
where ${\cal H}_{X_t; X_{t-1}}$ is the conditional entropy of $X_t$, given its recent past value $X_{t-1}$ [see Eq.~\eqref{eq:7}]. Moreover, ${\cal H}_{X_{t};X_{t-1},Y_{t-1}}$ denotes entropy of $X_t$, conditioned on the past values of both $X$ and $Y$, given by,
\begin{equation}
    {\cal H}_{X_{t};X_{t-1},Y_{t-1}} = S_{X_{t},X_{t-1},Y_{t-1}}-S_{X_{t-1},Y_{t-1}}, \label{eq:Hxyy}
\end{equation}
where $S_{X_{t-1},Y_{t-1}} = S_{X,Y}$ is the joint entropy of $X$ and $Y$, defined in Eq.~\eqref{eq:Sxy}, and
\begin{align}
&  S_{X_{t},X_{t-1},Y_{t-1}}  = -\int dx_t\int dx_{t-1}  \int dy_{t-1}\,\cr 
  & \quad \quad \times \, p(x_t,x_{t-1},y_{t-1})\,\ln p(x_t,x_{t-1},y_{t-1}), \label{eq:Sxyz}
\end{align}
denotes the joint entropy of $X$ with past values of itself and $Y$. Using Eqs.~\eqref{eq:Hxyy}-\eqref{eq:Sxyz}, Eq.~\eqref{eq:Txy} simplifies to,
\begin{align}
{\cal T}_{Y\to X}=S_{X_{t},X_{t-1}}-
 S_{X_{t},X_{t-1},Y_{t-1}}+S_{X,Y}-S_{X}, \label{eq:Tyx_2}
\end{align}
Note that, ${\cal T}_{Y\to X}$ is not symmetric in $X$ and $Y$. A positive ${\cal T}_{Y\to X}$ implies that past values of $Y$ has a directional effect on present values of $X$. This helps in identifying causal relationships, such as whether changes in $Y$ precede changes in $X$ or vice versa. 




We compute the transfer entropies  ${\cal T}_{\text{RH}\to \text{PM}_{2.5}}$ and ${\cal T}_{\text{PM}_{2.5}\to\text{RH}}$ using Eq.~\eqref{eq:Tyx_2}, for all the cities. 
Table~\ref{tab:TE} reports these values for the different groups of cities identified earlier on the basis of composite correlation index. Defining $\Delta {\cal T} = [{\cal T}_{\text{RH}\to \text{PM}_{2.5}}-\,\,{\cal T}_{\text{PM}_{2.5}\to\text{RH}}]$ allows us to interpret  $\Delta {\cal T}>0$ as RH leading PM\textsubscript{2.5} while $\Delta {\cal T}<0$ as PM\textsubscript{2.5} leading RH. 
It is evident from Table~\ref{tab:TE}, $\Delta {\cal T}>0$ for majority of cities, indicating  relative humidity (RH) predominantly exerts a forward influence on PM\textsubscript{2.5} concentrations. In contrast, a smaller subset of cities, namely, Delhi, Bhubaneswar, Guwahati, Mangaluru, Lucknow, Thiruvananthapuram, Visakhapatnam, and Hyderabad, exhibits the opposite ordering, where PM\textsubscript{2.5} variations precede changes in RH.

It is noteworthy that the magnitudes of $\Delta {\cal T}$ remain close to zero across most locations, with the sole exception of Srinagar ($\Delta {\cal T}$ = 0.272), likely reflecting its distinct meteorological regime due to geographic factors. All other values are substantially smaller, indicating limited directional asymmetry in information flow. The relatively small values of $\Delta {\cal T}$, irrespective of the correlation strength between PM\textsubscript{2.5} and RH, suggest that neither variable solely governs the underlying dynamics. Instead, the comparable TE magnitudes in both directions across cities indicate a bidirectional relationship between PM\textsubscript{2.5} and RH characterized by their mutual influence~\cite{Chen_2017}. This highlights the importance of considering RH and PM\textsubscript{2.5} as components of a coupled dynamical system in the study of air quality.

\begin{table}[th]
\centering
\begin{tabular}{|c|p{0.6 cm}|c|c|p{0.9 cm}|}
\hline 
$C$ & City & {${\cal T}_{\text{\tiny RH} \to \text{\tiny PM}_{2.5}}$} & {${\cal T}_{\text{\tiny PM}_{2.5}\to\text{\tiny RH}}$} & $\Delta {\cal T}$ \\
\hline
\multirow{4}{*}{} & AGT & 0.486 & 0.436 &\cellcolor{blue!25}0.050\\
Very & ASN &  0.738  & 0.660 &\cellcolor{blue!25}0.078\\
High& BDI &  0.411 & 0.404 &\cellcolor{blue!25}0.007\\
& BKN &  0.507 & 0.403 &\cellcolor{blue!25}0.104\\
\hline
\multirow{8}{*}{} & DEL & 0.038 & 0.046 &\cellcolor{red!25}-0.008\\
 & BGP &  0.333  & 0.311 &\cellcolor{blue!25}0.023\\
& BPL & 0.266 & 0.230 &\cellcolor{blue!25}0.036\\
High& BSR &  0.566  & 0.591 &\cellcolor{red!25}-0.025\\
& GWH & 0.192 & 0.210 &\cellcolor{red!25}-0.018\\
& IND &  0.267  & 0.235 &\cellcolor{blue!25}0.032\\
& KOP & 0.332  & 0.247 &\cellcolor{blue!25}0.085\\
& MAN & 0.119 & 0.198 &\cellcolor{red!25}-0.079\\
\hline
\multirow{7}{*}{} & CHN & 0.088&0.083 &\cellcolor{blue!25}0.005\\
& LKO & 0.129 &0.145 &\cellcolor{red!25}-0.016\\
Inter-& PDY &  0.382 &0.281 &\cellcolor{blue!25}0.101\\
mediate& RPR & 0.242 &0.185 &\cellcolor{blue!25}0.057\\
& SXR &  0.549 & 0.277 &\cellcolor{blue!25}0.272\\
& TVM &  0.147 &0.156 & \cellcolor{red!25}-0.009\\
& VTZ &  0.209 &0.257 &\cellcolor{red!25} -0.048\\
\hline
\multirow{5}{*}{} & AMD &  0.118 & 0.083 &\cellcolor{blue!25}0.035\\
&BLR & 0.028 & 0.021 & \cellcolor{blue!25}0.007\\
Low&HYD &  0.132  & 0.196 &\cellcolor{red!25} -0.064\\
& KOL & 0.093 & 0.072 & \cellcolor{blue!25}0.021\\
&MUM & 0.069 &  0.055 & \cellcolor{blue!25}0.014\\
\hline
\end{tabular}
\caption{Transfer Entropy ${\cal T}$ between PM\textsubscript{2.5} and RH pair across cities highlighting positive and negative $\Delta {\cal T}$ by blue and red color respectively.}
\label{tab:TE}
\end{table}

\begin{figure}[tbh]
    \centering \includegraphics[width=1\linewidth]{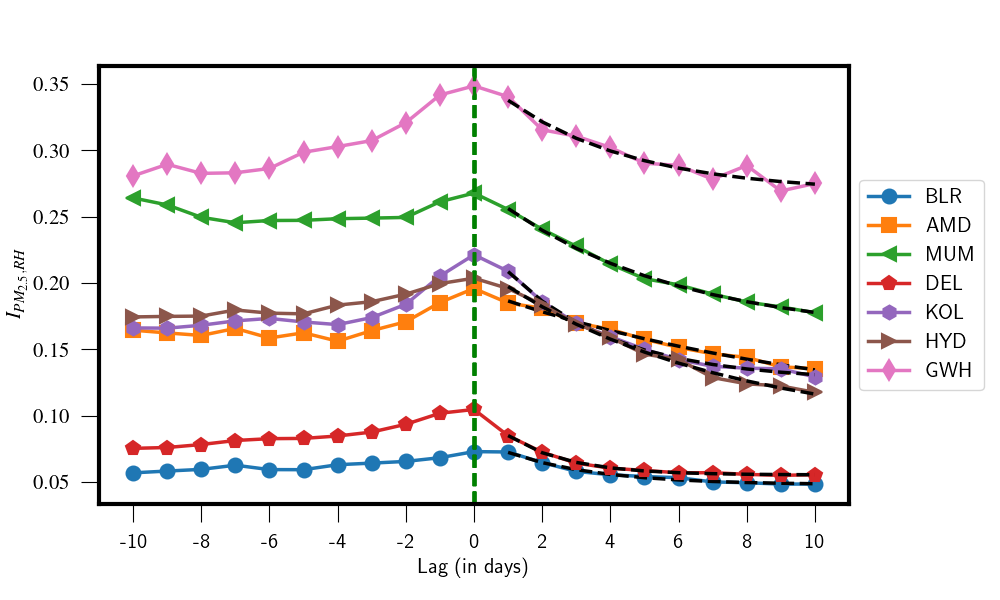} \includegraphics[width=1\linewidth]{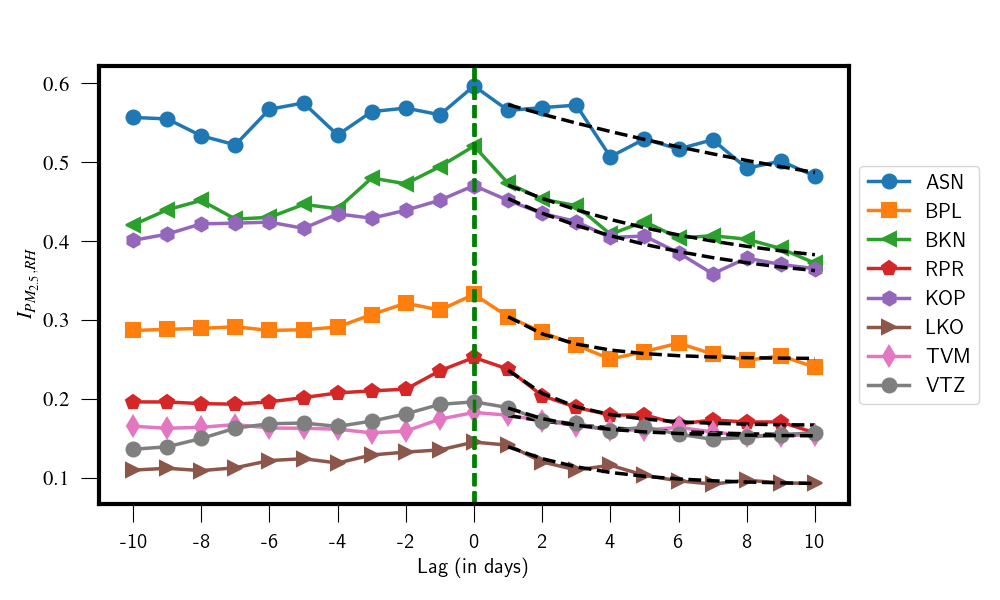}
    \caption{TDMI curves for PM\textsubscript{2.5}-RH pair across cities showing exponential decay with time-lag $\tau$.}
    \label{fig:TDMI_low}
\end{figure}

\subsection{Time-Delayed Mutual Information}
To further explore the temporal dependence between  PM\textsubscript{2.5} and RH, we compute the mutual information with a time delay, $\tau$, considering  PM\textsubscript{2.5} at day $t$ and RH at a later day $t +\tau$.
For two stationary time-series $X=\{x_t\}_{t=1}^N$ (PM\textsubscript{2.5}) and $Y=\{y_t\}_{t=1}^N$ (RH), the Time-Delayed Mutual Information (TDMI) for a given lag $\tau$ is defined as,
\begin{align}
I_{X,Y}(\tau) = \sum_{x_t}\sum_{y_{t+\tau}} p(x_t,y_{t+\tau})\,
\ln\left[\frac{p(x_t,y_{t+\tau})}{p(x)p(y)}\right],
\end{align}
where $p(x_t,y_{t+\tau})$ is the joint probability distribution of $(X_t,Y_{t+\tau})$ and $p(x)$, $p(y)$ are the corresponding marginals. Equivalently, in terms of entropies, we can write,
\begin{align}
I_{X,Y}(\tau) = S_{X} + S_{Y} - S_{X_t,Y_{t+\tau}},
\end{align}
with $S_{X_t,Y_{(t+\tau)}}$ denoting the joint entropy.
TDMI thus quantifies lagged statistical dependencies between PM\textsubscript{2.5} and RH. We evaluate $I_{X,Y}(\tau)$ for $\tau \in [-10,10]$ days, with results shown in Figure~\ref{fig:TDMI_low}. 

This analysis reveals how mutual information decays with time, offering insight into the response timescale. 
Figure~\ref{fig:TDMI_low} displays that in most of the cities,  TDMI curve attains its global maximum at zero lag ($\tau=0$) and then follows an approximately exponential decline as $\tau$ increases. This pattern indicates an almost instantaneous adjustment of PM$_{2.5}$ concentrations to humidity fluctuations, with the influence of any single RH perturbation fading rapidly, typically within a few days, suggesting limited memory in the underlying PM$_{2.5}$–RH interactions. 

The TDMI profiles for some of the cities differ markedly from the above behavior. The peak occurs at a finite, either positive or negative lag, and the subsequent decay deviates significantly from an exponential form. Such features imply a delayed and more persistent response of fine-particulate levels to humidity changes, possibly reflecting local factors.

\section{Conclusion}\label{sec:concl}
Our work contributes to the expanding application of statistical physics and information-theoretic tools to the study of real-world environmental systems.
Using an entropy-based dual framework, we analyze both static interdependencies and dynamic information flow to investigate correlations between air pollutants and meteorological variables across multiple Indian cities. 
By combining both linear and nonlinear dependency measures, namely, Pearson correlation, mutual information, relative conditional entropy, as well as  transfer entropy and time-delayed mutual information, we build a clearer picture of how environmental variables jointly evolve in urban atmospheres.

In the static analysis, a composite correlation index (CCI) is developed by amalgamating Pearson correlation, mutual information and relative conditional entropy. CCI  quantifies the interdependencies between the pollutants and meteorological variables 
and  helps to classify cities into distinct groups based on its strength.
High values of this index indicate strong, potentially nonlinear dependencies, while low values suggest weaker or more exogenous interactions. This classification has important implications for model selection and forecasting---regions with high CCI may benefit from nonlinear or machine learning models capable of capturing complex dependencies, while areas with low CCI might be adequately modeled using simpler, linear techniques.

On the other hand, TE and TDMI offer insights into the temporal structure of these dependencies. TE analysis of the PM$_{2.5}$–relative humidity (RH) pair shows bidirectional information flow in several cities, suggesting feedback mechanisms between pollutant levels and humidity. The TDMI results further indicate that, in many cases, mutual information peaks at lag zero and decays exponentially with increasing lag implying predominantly contemporaneous interactions with limited temporal memory. These decay profiles may serve as useful indicators of system responsiveness and reactivity.

This entropy-based framework opens multiple avenues for future research. 
Extending this  analysis to seasonal or diurnal cases may reveal important modulations in pollutant-weather interactions across various timescales. Employing multivariate transfer entropy can uncover higher-order dependencies and hidden intermediaries among multiple pollutants and weather drivers. 
Incorporating conditional or multivariate time-delayed mutual information  methods, along with additional meteorological parameters, could further enhance the analysis.
Finally, applying this framework across diverse climate-regions globally would test its generalizability and possibly uncover universal versus region-specific signatures in pollutant-weather interactions.

\section{Acknowledgment}
KG and MD  acknowledge  the support provided by the Indian Statistical Institute, Kolkata during this project.

 \section{Statements and Declarations}

\begin{itemize}
 \item \textbf{Data availability Statement}: All the data used in this work are publicly available~\cite{1}.
   \item \textbf{Competing Interests:} There are no competing interests to declare. 
   \item \textbf{Funding:} No funding information to declare.
   
 \end{itemize}

\begin{table*}[th]
    \centering
    \small%
    \begin{tabular}{|p{0.4 cm}|p{2.9 cm}|p{2.0 cm}|p{1.5 cm}|c|p{2 cm}|p{1.6 cm}|p{1.1 cm}|}
    \hline
    Sl. No. & City Name & State/ UT & Population & $N_m$ & Analysis period & Lat, Long & Climate zone\\ 
   \hline
   \hline 
      1& Ahmedabad (AMD)&Gujarat&90,61,820&3& June 2021 - Dec 2024
& 23.02\textdegree N, 72.57\textdegree E
& BSh
\\
         \hline
      2
& Bengaluru(BLR)&Karnataka&1,43,95,400&5& Jan 2020 - Dec 2024
& 12.97\textdegree N, 77.59\textdegree E
& Aw
\\
         \hline
      3& Chennai (CHN)&Tamil Nadu&1,23,36,000&7& Jan 2022 - Dec 2024& 13.08\textdegree N, 80.27\textdegree E
& As\\
         \hline
      4& Delhi (DEL)&Delhi NCR&3,46,65,600&23& Jan 2020 - Dec 2024
& 28.70\textdegree N, 77.10\textdegree E
& Cwa, BSh\\
         \hline
       5& Hyderabad (HYD)&Andhra Pradesh&1,13,37,900&5& Jan 2020 - Dec 2024& 17.39\textdegree N, 78.49\textdegree E
& BSh\\
         \hline
    6& Kolkata (KOL)&West Bengal&1,58,45,200&7& Jan 2020 - Dec 2024
& 22.57\textdegree N, 88.36\textdegree E
& Aw
\\
         \hline
     7& Mumbai (MUM)&Maharashtra&2,20,89,000 &8 & Jan 2020 - Dec 2024
& 19.08\textdegree N, 72.88\textdegree E
& Aw, Am\\
         \hline
      8& Agartala (AGT)& Tripura & 6,70,388&1& Nov 2020 - Dec 2024& 23.83\textdegree N, 91.29\textdegree E
& Aw\\
         \hline
    9& Asansol (ASN)&West Bengal&15,65,300&1& 
Sept 2022 - Dec 2024
& 23.68\textdegree N, 86.98\textdegree E
& Cwa\\
         \hline
      10& Baddi (BDI)&Himachal Pradesh&29,911&1& Mar 2022 - Dec 2024
& 30.90\textdegree N, 76.80\textdegree E
& Cwa
\\
         \hline
       11& Bhagalpur (BGP)&Bihar&5,25,429&2& 
Jan 2022 - Dec 2024
& 25.25\textdegree N, 86.99\textdegree E
& Cwa
\\
         \hline
       12& Bhopal (BPL)&Madhya Pradesh&26,86,290&3& Feb 2023 - Dec 2024
& 23.26\textdegree N, 77.41\textdegree E
& Cwa
\\
         \hline
          13& Bhubaneswar (BSR)&Odisha&13,20,910&2& Jan 2024 - Dec 2024
& 20.30\textdegree N, 85.82\textdegree E
& Aw
\\
         \hline
      14& Bikaner (BKN)&Rajasthan&8,53,869&1& Feb 2023 - Dec 2024
& 28.02\textdegree N, 73.31\textdegree E
& Bah\\
         \hline
      15& Guwahati (GWH)&Assam&12,24,170&1& Jan 2020 - Dec 2024
& 26.14\textdegree N, 91.74\textdegree E
& Cwa
\\
         \hline
      16& Indore (IND)&Madhya Pradesh&34,82,830&1& Jan 2022 - Dec 2024
& 22.72\textdegree N, 75.86\textdegree E
& Aw\\
         \hline
       17& Kolhapur (KOP)&Maharashtra&6,69,008&2& 
Apr 2023 - Dec 2024
& 16.70\textdegree N, 74.24\textdegree E
& Aw\\
         \hline
       18& Lucknow (LKO)&Uttar Pradesh&41,32,670&4& Jan 2022 - Dec 2024
& 26.85\textdegree N, 80.95\textdegree E
& Cwa
\\
         \hline
          19& Mangaluru (MAN)&Karnataka&7,78,685&1& 
Jul 2021 - Dec 2024
& 12.91\textdegree N, 74.86\textdegree E
& Am
\\
         \hline
    20& Puducherry (PDY)&Puducherry&9,40,911&1& Jan 2022 - Dec 2024
& 11.94\textdegree N, 79.81\textdegree E
& Aw
\\
         \hline
21& Raipur (RPR)&Chhattisgarh&19,23,440&4& Jan 2023 - Dec 2024
& 21.25\textdegree N, 81.63\textdegree E
& Aw
\\
         \hline
    22& Srinagar (SXR)&Jammu and Kashmir&17,77,610&1& 
Jan 2022 - Dec 2024
& 34.08\textdegree N, 74.80\textdegree E
& Cfa\\
         \hline
  23& Thiruvananthapuram (TVM)&Kerala&30,72,530&2& Jan 2020 - Dec 2024
& 8.52\textdegree N, 76.94\textdegree E
&Am
\\\hline
  24& Visakhapatnam (VTZ)&Andhra Pradesh&10,63,178&1& Jan 2020 - Dec 2024
& 17.69\textdegree N, 83.22\textdegree E
&Aw
\\\hline
    \end{tabular}
    \caption{Dataset summary and characteristics of the selected Indian cities.}
    \label{tab:1}
\end{table*}

\begin{table*}[th]
\centering
\small
\begin{minipage}{0.45\textwidth}
\centering
\begin{tabular}{|p{0.7 cm}||c|c|c|c|c|c|}
    \hline
    \textbf{City} & $S_{\text{PM}_{2.5}}$ & $S_{\text{PM}_{10}}$ & $S_{\text{SO}_{2}}$ & $S_{\text{NO}_{2}}$ & $S_{\text{RH}}$ & $S_{\text{AT}}$ \\
    \hline
    AMD & 4.533 & 5.259 & 3.708 & 4.149 & \cellcolor{blue!25}4.394 & 2.941 \\
    \hline
    BLR & 4.167 & 4.888 & 2.477 & 3.801 & 3.987 & 2.205 \\
    \hline
    CHN & 4.204 & 4.849 & 2.854 & 3.696 & 3.531 & 2.350 \\
    \hline
    DEL & \cellcolor{blue!25}6.158 & 5.603 & 3.861 & \cellcolor{blue!25}4.459 & 4.066 & \cellcolor{blue!25}3.338 \\
    \hline
    HYD & 4.336 & 5.120 & 2.932 & 4.215 & 4.039 & 2.695 \\
    \hline
    KOL & 4.745 & 5.368 & 2.951 & 4.083 & 3.925 & 2.799 \\
    \hline
    MUM & 4.465 & 5.399 & 3.629 & 4.016 & 3.860 & 2.636 \\
    \hline
    AGT & 4.631 & 5.091 & 3.835 & 3.128 & 3.784 & 3.163 \\
    \hline
    ASN & 4.866 & 5.607 & 2.739 & 3.591 & 3.927 & 2.994 \\
    \hline
    BDI & 4.809 & 5.509 & \cellcolor{blue!25}3.950 & 3.465 & 4.271 & 3.065 \\
    \hline
    BGP & 5.178 & \cellcolor{blue!25}5.825 & 3.573 & 4.337 & 3.998 & 3.264 \\
    \hline
    BPL & 4.565 & 5.246 & 3.360 & 3.884 & 4.287 & 3.047 \\
    \hline
\end{tabular}
\end{minipage}%
\hfill
\begin{minipage}{0.45\textwidth}
\centering
\begin{tabular}{|p{0.7 cm}||c|c|c|c|c|c|}
    \hline
    \textbf{City} & $S_{\text{PM}_{2.5}}$ & $S_{\text{PM}_{10}}$ & $S_{\text{SO}_{2}}$ & $S_{\text{NO}_{2}}$ & $S_{\text{RH}}$ & $S_{\text{AT}}$ \\
    \hline
    BSR & 4.547 & 5.217 & 3.002 & 3.335 & 3.903 & 2.628 \\
    \hline
    BKN & 4.739 & 5.782 & 2.133 & 3.956 & 4.201 & 3.231 \\
    \hline
    GWH & 4.937 & 5.492 & 3.380 & \cellcolor{red!25}2.677 & 3.918 & 2.873 \\
    \hline
    IND & 4.397 & 5.256 & 2.856 & 4.506 & 4.253 & 2.925 \\
    \hline
    KOP & 4.487 & 5.227 & \cellcolor{red!25}1.776 & 3.762 & 3.622 & 2.109 \\
    \hline
    LKO & 4.680 & 5.378 & 3.258 & 3.658 & 4.207 & 3.285 \\
    \hline
    MAN & 4.021 & 4.739 & 3.241 & 3.745 & \cellcolor{red!25}3.113 & \cellcolor{red!25}0.546 \\
    \hline
    PDY & \cellcolor{red!25}3.911 & \cellcolor{red!25}4.430 & 2.739 & 2.892 & 3.202 & 2.235 \\
    \hline
    RPR & 4.209 & 5.027 & 2.878 & 4.219 & 4.291 & 2.953 \\
    \hline
    SXR & 3.918 & 4.967 & 3.561 & 3.273 & 3.906 & 3.248 \\
    \hline
    TVM & 3.972 & 4.490 & 2.462 & 3.176 & 3.670 & 2.217 \\
    \hline
    VTZ & 4.558 & 5.301 & 2.922 & 3.908 & 3.364 & 2.688 \\
    \hline
\end{tabular}
\end{minipage}
\caption{Self-entropy values of the selected parameters across cities highlighting maximum (blue) and minimum (red) values of each variables.}
\label{tab:Entropy}
\end{table*}
\begin{table*}[th]
\centering
\small
 \begin{tabular}{|c||c|c|c|c|c|c|c|c|}
\hline
{\textbf{City}} & ${\cal H}_{\text{PM}_{2.5};\text{RH}}$ & ${\cal H}_{\text{PM}_{2.5};\text{AT}}$ & ${\cal H}_{\text{PM}_{10};\text{RH}}$ & ${\cal H}_{\text{PM}_{10};\text{AT}}$ & ${\cal H}_{\text{SO}_{2};\text{RH}}$ & ${\cal H}_{\text{SO}_{2};\text{AT}}$ & ${\cal H}_{\text{NO}_{2};\text{RH}}$ & ${\cal H}_{\text{NO}_{2};\text{AT}}$\\
\hline
AMD
& 4.364& 4.476& 5.083& 5.204& 3.301& 3.476& 3.696& 3.884
\\\hline
BLR
& 4.122& 4.212& 4.771& 4.847& 2.411& 2.471& 3.702&3.765
\\\hline
 CHN
& 4.001& 4.033& 4.706& 4.713& 2.805& 2.811& 3.522&3.582
\\\hline
DEL
& \cellcolor{blue!25}5.464& \cellcolor{blue!25}5.335& \cellcolor{blue!25}6.052& \cellcolor{blue!25}6.016& 3.240& 3.283& \cellcolor{blue!25}4.502& \cellcolor{blue!25}4.452
\\
    \hline
    HYD
& 4.142& 4.264& 4.890& 5.045& 2.867& 2.836& 4.096& 4.119
\\
    \hline
KOL
& 4.610& 4.336& 5.213& 4.948& 2.936& 2.784& 4.054& 3.848
\\
    \hline
MUM
& 4.278& 4.393& 5.160& 5.347& \cellcolor{blue!25}3.639& 3.585& 3.915& 3.960
\\\hline
 AGT
& 4.244& 4.265& 4.796& 4.760& 3.620& 3.602& 2.960&2.945
\\\hline
 ASN
& 4.158& 4.524& 4.812& 5.233& 2.297& 2.604& 3.045&3.289
\\
    \hline
BDI
& 4.339& 4.449& 5.158& 5.356& 3.612& \cellcolor{blue!25}3.762& 3.269& 3.216
\\
    \hline
BGP
&  4.836&  4.713&  5.595&  5.601&  3.459&  3.474&  3.923& 3.962
\\
    \hline
    
BPL
& 4.259& 4.224& 4.950& 5.044& 3.273& 3.321& 3.650& 3.747
\\
    \hline    
BSR
&  4.173&  3.917&  4.784&  4.637&  2.738&  2.677&  2.825& 2.555
\\
    \hline
BKN
&  4.248&  4.235&  5.448&  5.621&  1.917&  1.970&  3.724& 3.803
\\
    \hline
    
GWH
&  4.559&  4.454&  5.124&  5.058&  2.883&  2.946&  \cellcolor{red!25}1.942& \cellcolor{red!25}1.728
\\
    \hline
IND
&  4.007&  4.205&  4.801&  5.128&  2.56&  2.830&  4.059& 4.407
\\
    \hline
    
KOP
&  3.945&  4.355&  4.656&  5.112&  \cellcolor{red!25}1.025&  \cellcolor{red!25}1.471&  3.284& 3.660
\\
    \hline
    LKO
& 4.468& 4.527& 5.106& 5.246& 3.121& 3.215& 3.339& 3.473
\\
    \hline
 MAN
& 3.909& 3.980& 4.641& 4.673& 3.040& 3.108& 3.313&3.504
\\\hline
 PDY
& \cellcolor{red!25}3.771& 3.731& \cellcolor{red!25}4.319& 4.359& 2.702& 2.709& 2.852&2.707
\\\hline
 RPR
& 3.930& 4.082& 4.756& 4.929& 2.618& 2.702& 3.741&3.978
\\ \hline
SXR
& 3.860& 3.795& 4.871& 4.792& 3.406& 3.382& 3.104&3.021
\\\hline
 TVM
& 3.797& \cellcolor{red!25}3.695& 4.347& \cellcolor{red!25}4.217& 2.161& 2.003& 3.052&2.994
\\\hline
VTZ
& 4.305& 4.191& 5.056& 4.981& 2.700& 2.683& 3.710&3.637
\\\hline
    \end{tabular}
\caption{City-wise conditional entropy values of air pollutants with respect to meteorological parameters highlighting maximum (blue) and minimum (red) values.}
    \label{tab:Cond_Entropy}
\end{table*}

\begin{table*}[th]
    \centering
    \small
    \begin{tabular}{|c||c|c|c|c|c|c|c|c|}
\hline
{\textbf{City}} & $C_{\text{\tiny PM}_{2.5},\text{\tiny RH}}$ & $C_{\text{\tiny PM}_{2.5},\text{\tiny AT}}$ & $C_{\text{\tiny PM}_{10},\text{\tiny RH}}$ & $C_{\text{\tiny PM}_{10},\text{\tiny AT}}$ & $C_{\text{\tiny SO}_{2},\text{\tiny RH}}$ & $C_{\text{\tiny SO}_{2},\text{\tiny AT}}$ & $C_{\text{\tiny NO}_{2},\text{\tiny RH}}$ & $C_{\text{\tiny NO}_{2},\text{\tiny AT}}$\\
\hline
AMD& 0.265&  0.270&  0.394&  0.344&  0.390&  0.399&  0.458&  0.341
\\
    \hline
    
BLR& 0.201&  0.236&  0.340&  0.370&  0.288&  0.234&  0.278&  0.228
\\
    \hline
    CHN& 0.427&  0.275&  0.397&  0.325&  0.280&  0.227&  0.394&  0.256
\\
    \hline
    
DEL& 0.507&  0.515&  0.353&  0.052&  0.301&  0.437&  0.248&  0.193
\\
    \hline
    HYD& 0.263&  0.256&  0.358&  0.332&  0.304&  0.402&  0.275&  0.212
\\
    \hline
    
KOL& 0.184&  0.375&  0.314&  0.409&  0.099&  0.280&  0.222&  0.202
\\
    \hline
    MUM& 0.201& 0.232& 0.328& 0.314&0.211& 0.284&0.196&0.222
\\
   \hline
    
AGT&0.718& 0.569& 0.624& 0.616& 0.342&0.436&0.597&0.453
\\\hline
 ASN&0.757& 0.503& 0.713& 0.544& 0.462& 0.495& 0.546&0.380
\\
    \hline
    BDI& 0.702& 0.442& 0.579& 0.454&0.531& 0.427&0.614&0.324
\\
    \hline
    
BGP& 0.528& 0.410& 0.459& 0.385&0.344& 0.360&0.446&0.345
\\
    \hline
    BPL& 0.479& 0.392& 0.476& 0.411&0.291& 0.385&0.389&0.288
\\
    \hline
BSR& 0.573& 0.670& 0.598& 0.667&0.318& 0.556&0.660&0.552
\\
        \hline
    BKN& 0.797& 0.545& 0.631& 0.512&0.371& 0.417&0.524&0.367
\\
        \hline
    GWH& 0.572& 0.506& 0.607& 0.546&0.733& 0.725&0.759&0.600
\\
        \hline
    IND& 0.472& 0.340& 0.502& 0.459&0.234& 0.352&0.388&0.317
\\
        \hline
    KOP
& 0.543& 0.397& 0.573& 0.519&0.588& 0.570&0.494&0.368
\\
        \hline
    LKO
& 0.342& 0.264& 0.400& 0.323&0.384& 0.363&0.357&0.248
\\
        \hline
    MAN
& 0.516& 0.364& 0.627& 0.468&0.414& 0.585&0.373&0.515
\\
        \hline
    PDY
& 0.413& 0.339& 0.423& 0.418&0.265& 0.427&0.418&0.270
\\
        \hline
    RPR
& 0.341& 0.296& 0.432& 0.379&0.247& 0.542&0.359&0.305
\\
        \hline
    SXR
& 0.327& 0.434& 0.508& 0.489&0.300& 0.422&0.518&0.412
\\
        \hline
    TVM
& 0.390& 0.622& 0.465& 0.685&0.282& 0.562&0.448&0.391
\\
        \hline
 VTZ
& 0.370& 0.401& 0.468& 0.469& 0.418& 0.432& 0.366&0.344
\\ \hline 
    \end{tabular}
    \caption{Composite correlation index ($C$) between atmospheric pollutants and meteorological variable pairs across cities.}
    \label{tab:comp}
\end{table*}

\bibliography{ref}

\end{document}